\documentclass[floatfix, print, twocolumn, prd,
showpacs, showkeys, preprintnumbers, nofootinbib, superscriptaddress]{revtex4-1}
\usepackage{natbib}
\usepackage{times}
\usepackage{amssymb,amsbsy,amsmath,amsfonts}
\usepackage{graphicx}
\usepackage{float}
\usepackage{color}
\usepackage{morefloats}
\usepackage{rotating}
\usepackage{srcltx}
\usepackage{slashed}
\usepackage{subfigure}
\usepackage{multirow}
\usepackage{verbatim}
\usepackage{hyperref}
\usepackage{tabularx}
\usepackage{epstopdf}

\graphicspath{{./}{./img/}{./fig/}{./image/}{./figure/}{./picture/}}

\begin{document}

\title{Scalar strangeness content of the nucleon and baryon sigma terms
}

\author{Xiu-Lei Ren}
\affiliation{School of Physics and
Nuclear Energy Engineering and International Research Center for Nuclei and Particles in the Cosmos, Beihang University, Beijing 100191, China}

\author{Li-Sheng Geng}
\email[E-mail: ]{lisheng.geng@buaa.edu.cn}
\affiliation{School of Physics and
Nuclear Energy Engineering and International Research Center for Nuclei and Particles in the Cosmos, Beihang University, Beijing 100191, China}

\author{Jie Meng}
\affiliation{School of Physics and
Nuclear Energy Engineering and International Research Center for Nuclei and Particles in the Cosmos, Beihang University, Beijing 100191, China}
\affiliation{State Key Laboratory of Nuclear Physics and Technology, School of Physics, Peking University, Beijing 100871, China}
\affiliation{Department of Physics, University of Stellenbosch, Stellenbosch 7602, South Africa}

\begin{abstract}
The scalar strangeness content of the nucleon, characterized by the so-called strangeness-nucleon  sigma term,  is of fundamental importance in understanding its sea-quark flavor structure. We report a determination of the octet baryon sigma terms via the Feynman-Hellmann theorem by analyzing the latest high-statistics $n_f=2+1$ lattice QCD simulations with
covariant baryon chiral perturbation theory up to next-to-next-to-next-to-leading order.  In particular, we predict $\sigma_{\pi N}=55(1)(4)$ MeV and $\sigma_{sN}=27(27)(4)$ MeV, while the first error is statistical and the second systematic due to  different lattice scales. The predicted $\sigma_{sN}$ is consistent with the latest LQCD results
and the results based on the next-to-next-to-leading order chiral perturbation theory. Several key factors in determining the sigma terms are systematically taken into account and clarified for the first time, including 
the effects of lattice scale setting,   systematic uncertainties originating from chiral expansion truncations, and constraint of strong-interaction isospin breaking effects. 
\end{abstract}

\pacs{12.39.Fe,  12.38.Gc, 14.20.Dh, 14.20.Jn}
\keywords{Chiral Lagrangians, Lattice QCD calculations, Protons and neutrons, Hyperons}

\date{\today}

\maketitle

\section{Introduction}
Understanding  the sea-quark structure  of the nucleon has
long been a central topic in nuclear physics~\cite{Chang:2014jba}.  Of particular interest are the contributions of the $s\bar{s}$ component since the nucleon contains no valence strange quarks, e.g., the strangeness contribution to the proton spin~\cite{QCDSF:2011aa} and 
to the electric and magnetic form factors~\cite{Ahmed:2011vp}. In this context, the strangeness-nucleon sigma term, $\sigma_{sN}=m_s\langle N|\bar{s}s|N\rangle$, plays an important role as it
relates to the scalar strangeness content of the nucleon,  the composition of the nucleon mass,  $K N$ scatterings,  counting rates in Higgs boson searches~\cite{Cheng:1988cz}, and the precise measurement of the standard model parameters in $pp$ collisions at the LHC~\cite{Aad:2009wy}. Furthermore, the uncertainty in $\sigma_{sN}$  is
 the principal source in predicting the cross section of certain candidate dark matter particles interacting with the nucleons~\cite{Bottino:1999ei}.

Although  the pion-nucleon sigma term $\sigma_{\pi N}=m_l\langle N|\bar{u}u+\bar{d}d|N\rangle$ can
 be determined from pion-nucleon scattering ~\cite{Gasser:1990ce,Alarcon:2011zs}, historically the strangeness-nucleon sigma term has been determined indirectly via  the nonsinglet matrix element $\sigma_0=m_l\langle N|\bar{u}u+\bar{d}d-2\bar{s}s|N\rangle$, which yields a value  ranging from 0 to 300 MeV~\cite{Alarcon:2012nr} .  In principle, lattice quantum chromodynamics (LQCD) provides  a model independent way in the determination of the baryon sigma terms by either computing three-point ~\cite{Takeda:2010cw,Babich:2010at,Bali:2011ks,Dinter:2012tt,Engelhardt:2012gd,Oksuzian:2012rzb,
Gong:2013vja,Alexandrou:2013nda} or two-point correlation functions (the so-called spectrum method)~\cite{Young:2009zb,Oksuzian:2012rzb,Durr:2011mp,Horsley:2011wr,Semke:2012gs,Shanahan:2012wh,Ren:2012aj,Jung:2012rz,Junnarkar:2013ac}. Although
tremendous efforts have been made in this endeavor , due to the many systematic and statistical uncertainties inherent in these studies, no consensus has been reached on
the value of the scalar strangeness content of the nucleon.
 
 The most important sources of systematic uncertainties originate from the so-called chiral extrapolations.  In the $u$, $d$, and $s$ flavor sector, a
 proper formulation of baryon chiral perturbation theory (BChPT) that satisfies all symmetry and analyticity constraints is known to be essential to properly describe the nonperturbative regime of QCD. In this sense, the extended-on-mass-shell (EOMS) formulation~\cite{Gegelia:1999gf}
 has shown a number of both formal and practical advantages, whose applications have solved a number of long-existing puzzles in the one-baryon sector~\cite{Geng:2013xn}. Its applications in the studies of the LQCD octet baryon masses
 turn out to be very successful as well~\cite{Ren:2012aj,Ren:2013dzt,Ren:2013oaa}.  Furthermore, as demonstrated recently, mass dependent and mass independent  lattice scale setting methods can result in a $\sigma_{sN}$ different by a factor of  three~\cite{Shanahan:2012wh,Shanahan:2013cd}, Therefore, it casts doubts on the determination of
 the $\sigma_{sN}$ from a single data set with a particular scale setting method. To say the least, systematic uncertainties might be well underestimated.

 In this article, utilizing the latest and high-statistics $n_f=2+1$ LQCD simulations of the octet   baryon masses from the PACS-CS
~\cite{Aoki:2008sm}, LHPC~\cite{WalkerLoud:2008bp}, and QCDSF-UKQCD~\cite{Bietenholz:2011qq} collaborations  and
  the Feynman-Hellmann theorem, paying special attention to the lattice scale setting, we report a  determination of the baryon sigma terms, particularly the strangeness-nucleon sigma term, in covariant BChPT up to 
  next-to-next-to-next-to-leading order (N$^3$LO).

\section{Feynman-Hellmann theorem and octet baryon sigma terms}
The Feynman-Hellmann theorem~\cite{Feynman:1939zza} dictates that at the isospin limit the baryon sigma terms can be calculated from the quark mass dependence of the
octet baryon masses, $m_B$, in the following way: 
\begin{eqnarray}
  \sigma_{\pi B} &=& m_l\langle B|\bar{u}u+\bar{d}d|B\rangle \equiv m_l\frac{\partial m_B}{\partial m_l},\\ 
  \sigma_{s B} &=& m_s\langle B|\bar{s}s|B\rangle \equiv m_s \frac{\partial m_B}{\partial m_s}.
\end{eqnarray}
Up to $\mathcal{O}(p^4)$ in the chiral expansion,
\begin{equation}\label{Eq:MB} 
  m_B = m_0 + m_B^{(2)} + m_B^{(3)} + m_B^{(4)},
\end{equation}
where $m_0$ is the SU(3) chiral limit octet baryon mass, and $m_B^{(2)}$, $m_B^{(3)}$, and $m_B^{(4)}$ represent the $\mathcal{O}(p^2)$, $\mathcal{O}(p^3)$, and $\mathcal{O}(p^4)$ chiral contributions~\cite{Ren:2012aj}. The virtual decuplet contributions are not explicitly included in Eq.~(\ref{Eq:MB}), since their effects on the chiral extrapolation and finite-volume corrections (FVCs) are shown to be relatively small~\cite{Ren:2013dzt}. 

At N$^3$LO, there are $19$ unknown low energy constants (LECs) ($m_0$, $b_0$, $b_D$, $b_F$, $b_{1-8}$, $d_{1-5,7,8}$) in Eq.~\eqref{Eq:MB} to be determined by fitting the LQCD data~\cite{Ren:2012aj}, and the others are fixed at the following values: $D=0.8$, $F=0.46$~\cite{Cabibbo:2003cu}, $F_\phi=0.0871$ GeV~\cite{Amoros:2001cp}, and $\mu=1$ GeV.

It should be stressed that in the present work the LQCD baryon masses only serve as inputs to fix the relevant LECs of the BChPT. Once the LECs are known, BChPT completely determines the derivatives of the baryon masses with respect to the light $u/d$ (strange) quark mass with the strange (light) quark mass fixed, thus yielding the desired baryon sigma terms.

\section{Three Key factors in an accurate determination of baryon sigma terms}
In order to obtain an accurate determination of the baryon sigma terms, a careful examination of the LQCD data is essential, since not all of them are of the same quality  though they are largely consistent with each other as shown in Refs.~\cite{Ren:2012aj,Lutz:2014oxa}. For instance, the statistics of the HSC simulations needs to be improved~\cite{Lin:2008pr} while the NPLQCD simulations are performed at one single combination of 
light-quark and strange-quark masses~\cite{Beane:2011pc}, which offers little constraint on the quark mass dependence of the baryon masses. The BMW simulations~\cite{Durr:2008zz}, though of high quality, are not publicly available. This leaves the PACS-CS
~\cite{Aoki:2008sm}, LHPC~\cite{WalkerLoud:2008bp}, and QCDSF-UKQCD~\cite{Bietenholz:2011qq} data for our study. It is important to note that most LQCD simulations fix the strange-quark mass close to its physical value and vary the light-quark masses. As a result, they  are suitable to study the light-quark mass dependence but not the strange-quark mass dependence. In this respect, the  QCDSF-UKQCD simulations are of particular importance because they provide a dependence of the baryon masses on the
strange-quark mass in a region not accessible in other simulations.  In order to stay within the application region of 
the BChPT, we only choose the LQCD data satisfying the following two criteria: $M_\pi<500$ MeV~\footnote{We have checked that
reducing the cut on $M_\pi$ down to $M_\pi=400 $ MeV or $M_\pi=360$ MeV has little effect on our numerical results, but since our $\chi^2/\mathrm{d.o.f.}$ (see Table I) is already about 1, there
is no need to further decrease the cut on $M_\pi$.} and $M_\phi L>3.8$,  as in Refs.~\cite{Ren:2013dzt,Ren:2013wxa}. It should be noted that the later criterium is relaxed for the
QCDSF-UKQCD data \footnote{The smallest  $M_\phi L$ taken into account is 2.932.} since their FVCs are small  because of the use of the ratio method~\cite{Bietenholz:2011qq}.  

A second issue relates to the scale setting of LQCD simulations (for a recent review, see Ref.~\cite{Sommer:2014mea}). For the spectrum determination of the baryon sigma terms, it was pointed out in Ref.~\cite{Shanahan:2012wh,Shanahan:2013cd} that using the  Sommer scale $r_0$~\cite{Sommer:1993ce} to fix the lattice spacing of the PACS-CS data can change the prediction of the strangeness-nucleon sigma term by a factor of two to three in
the finite-range regularization  (FRR) BChPT  up to next-to-next-to-leading order (NNLO).  It was claimed that using  $r_1$ for the purpose of lattice scale setting  is preferred in the LHPC simulations~\cite{WalkerLoud:2011ab} as well. As a result, it is necessary to understand how different scale setting methods for the same simulations affect the prediction of the baryon sigma terms.  Unfortunately, such a systematic study is still missing. In addition, instead of relying on the scale determined by the LQCD collaborations themselves, one can fix the lattice scale self-consistently in the BChPT study of the LQCD dimensionless data, as recently done in Ref.~\cite{Lutz:2014oxa}.  In the present work, all the three alternative ways of lattice scale setting will be studied and their effects on the predicted sigma terms examined and quantified.

\begin{table}[b]
  \centering
  \caption{Values of the LECs from the best fits to the LQCD data and the experimental octet baryon masses up to N$^3$LO. The lattice scale in each simulation
  is determined  using both the mass independent scale setting (MIS) and the MDS methods. In the MIS,
both the original lattice spacings determined by the LQCD collaborations themselves ``$a$ fixed'' and the self-consistently determined lattice spacings ``$a$ free'' are used (see text for details).}
  \label{Tab:N3LOLECs}
  \begin{tabular}{cccc}
  \hline\hline
    &  \multicolumn{2}{c}{MIS}  & MDS \\
  \cline{2-3}
    &  $a$ fixed & $a$ free& \\
  \hline
  $m_0$ [MeV]        &  $884(11)$    & $877(10)$    &  $887(10)$   \\
  $b_0$ [GeV$^{-1}$] &  $-0.998(2)$  & $-0.967(6)$  & $-0.911(10)$\\
  $b_D$ [GeV$^{-1}$] &  $0.179(5)$   & $0.188(7)$   & $0.039(15)$ \\
  $b_F$ [GeV$^{-1}$] &  $-0.390(17)$ & $-0.367(21)$  & $-0.343(37)$ \\
  \hline
  $b_1$ [GeV$^{-1}$] & $0.351(9)$    & $ 0.348(4)$  & $-0.070(23)$ \\
  $b_2$ [GeV$^{-1}$] & $0.582(55)$   & $ 0.486(11)$  & $0.567(75)$    \\
   $b_3$ [GeV$^{-1}$] & $-0.827(107)$ & $-0.699(169)$  & $-0.553(214)$  \\
  $b_4$ [GeV$^{-1}$] & $-0.732(27)$  & $-0.966(8)$  & $-1.30(4)$     \\
  $b_5$ [GeV$^{-2}$] & $-0.476(30)$  & $-0.347(17)$  & $-0.513(89)$   \\
  $b_6$ [GeV$^{-2}$] & $0.165(158)$  & $0.166(173)$   & $-0.0397(1574)$\\
  $b_7$ [GeV$^{-2}$] & $-1.10(11)$   & $-0.915(26)$  & $-1.27(8)$     \\
  $b_8$ [GeV$^{-2}$] & $-1.84(4)$    & $-1.13(7)$   & $0.192(30)$    \\
  \hline
  $d_1$ [GeV$^{-3}$] & $0.0327(79)$  & $0.0314(72)$  & $0.0623(116)$  \\
  $d_2$ [GeV$^{-3}$] & $0.313(26)$   & $0.269(42)$   & $0.325(54)$    \\
  $d_3$ [GeV$^{-3}$] & $-0.0346(87)$ & $-0.0199(81)$ & $-0.0879(136)$ \\
  $d_4$ [GeV$^{-3}$] & $0.271(30)$   & $0.230(24)$   & $0.365(23)$    \\
  $d_5$ [GeV$^{-3}$] & $-0.350(28)$  & $-0.302(50)$  & $-0.326(66)$   \\
  $d_7$ [GeV$^{-3}$] & $-0.435(10)$  & $-0.352(8)$  & $-0.322(7)$    \\
  $d_8$ [GeV$^{-3}$] & $-0.566(24)$  & $-0.456(30)$  & $-0.459(33)$   \\
  \hline
  $\chi^2/\mathrm{d.o.f.}$ & $0.87$  & $0.88$    & $0.53$   \\
  \hline\hline
\end{tabular}
\end{table}

\begin{figure*}[t]
  \centering
  \includegraphics[width=18cm]{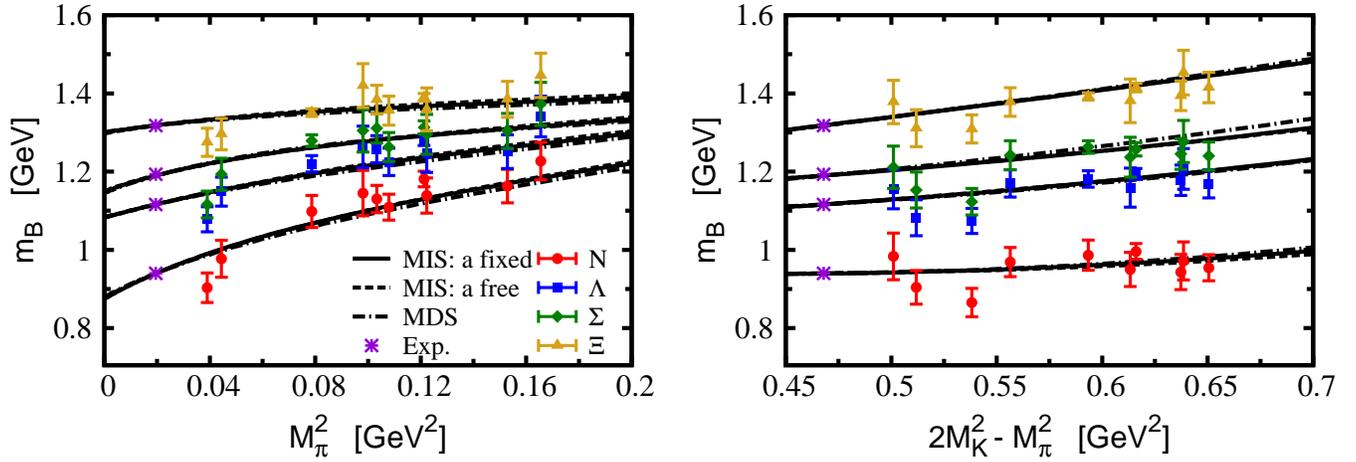}
  \caption{(color online). Octet baryon masses as a function of $M_\pi^2$ and  $2 M_K^2-M_\pi^2$ vs the BMW LQCD data~\cite{Durr:2011mp}. 
  The solid, dashed, and dot-dashed lines are obtained with the LECs from the three fits of Table ~\ref{Tab:N3LOLECs}.  
  On the left and right panels, the strange-quark mass and the light-quark mass are
  fixed at their respective physical values.  }
  \label{Fig:BMW}
\end{figure*}

At N$^3$LO, the large number of unknown LECs should be better constrained in order to give a reliable prediction of the baryon sigma terms. 
In this work, we employ the latest LQCD results on the strong isospin-splitting effects on the octet baryon masses to further constrain the LECs. The following values are used: $\delta m_N=-2.50(50)$ MeV, $\delta m_\Sigma=-7.67(79)(105)$ MeV and $\delta m_\Xi = -5.87(76)(43)$ MeV at the physical point.
  The $\delta m_N$ is chosen such as to cover all the recent results~\cite{Beane:2006fk,Borsanyi:2013lga},
  while the $\delta m_\Sigma$ and $\delta m_\Xi$ are taken from Ref.~\cite{Borsanyi:2013lga}. It should be pointed out that no new unknown LECs need to be introduced to calculate isospin-breaking corrections up to the order at which we work.

\begin{table}[t]
  \centering
  \caption{Predicted pion- and strangeness-sigma terms of the octet baryons  (in units of MeV) by the N$^3$LO BChPT with the LECs of Table \ref{Tab:N3LOLECs}.}
  \label{Tab:N3LOsigma}
\begin{tabular}{ccccc}
  \hline\hline
    & \multicolumn{2}{c}{MIS} & MDS \\
  \cline{2-3}
      & $a$ fixed & $a$ free &  \\
  \hline
  $\sigma_{\pi N}$     & $55(1)(4)$ & $54(1)$ & $51(2)$ \\
  $\sigma_{\pi \Lambda}$  & $32(1)(2)$ & $32(1)$ & $30(2)$  \\
  $\sigma_{\pi \Sigma}$    & $34(1)(3)$ & $33(1)$ & $37(2)$  \\
  $\sigma_{\pi \Xi}$      & $16(1)(2)$ & $18(2)$ & $15(3)$  \\
  \hline
  $\sigma_{s N}$        & $27(27)(4)$  & $23(19)$  & $26(21)$  \\
  $\sigma_{s \Lambda}$  & $185(24)(17)$ & $192(15)$ & $168(14)$ \\
  $\sigma_{s \Sigma}$  & $210(26)(42)$ & $216(16)$ & $252(15)$ \\
  $\sigma_{s \Xi}$     & $333(25)(13)$ & $346(15)$ & $340(13)$ \\
  \hline\hline
\end{tabular}
\end{table}

\section{BChPT study of LQCD octet baryon masses}
At N$^3$LO, the LQCD and experimental meson masses are described by the next-to-leading order ChPT~\cite{Gasser:1984gg} with the LECs of Ref.~\cite{Bijnes:1995AA}. FVCs~\cite{Hasenfratz:1989pk} are
taken into account but found to play a negligible role. In Table~\ref{Tab:N3LOLECs},  we tabulate the LECs and the corresponding $\chi^2/\mathrm{d.o.f.}$ from three best fits to the LQCD mass data and the experimental octet
baryon masses.
 In the first fit, we use the lattice spacings $a$ determined by the LQCD collaborations themselves to obtain the hadron masses in physical units as done in
 Ref.~\cite{Ren:2012aj}. In the second fit, we  determine the lattice spacings $a$ self-consistently.  Interestingly, we find that the so-determined lattice spacings $a$ are close to the ones determined by the LQCD collaborations themselves.   Specifically, the PACS-CS deviation is  2.5\%, the LHPC deviation is 4.1\%, and the 
 QCDSF-UKQCD deviation is 2.1\%. The corresponding $\chi^2/\mathrm{d.o.f.}$ also look similar. 
 While in the third fit, we adopt the so-called mass dependent scale setting (MDS), either from $r_0$ for the PACS-CS data  
 with $r_0(\mathrm{phys.})=0.465(12)$ fm~\cite{Aubin:2004wf}, $r_1$ for the LHPC data with $r_1(\mathrm{phys.})=0.31174(20)$ fm~\cite{Junnarkar:2013ac}~\footnote{Technically, this scale setting should be
 classified as a mass independent scale setting. Here, we slightly misuse the terminology to distinguish it from the one used in the LHPC original publication~\cite{WalkerLoud:2008bp}.}, 
 or $X_\pi$ for the QCDSF-UKQCD data with $X_\pi(\mathrm{phys.})=0.4109$ GeV~\cite{Bietenholz:2011qq}. The third fit yields a smaller $\chi^2/\mathrm{d.o.f.}$ and different LECs compared to the other two fits.

In Fig.~\ref{Fig:BMW}, we show the octet baryon masses as functions of $M_\pi^2$ ($2M_K^2-M_\pi^2$) using the LECs from Table~\ref{Tab:N3LOLECs} with the physical light- (right panel) and strange-quark  (left panel) masses. In order to cross-check the validity of our N$^3$LO BChPT fit, the BMW Collaboration data~\cite{Durr:2011mp} are shown as well. It is clear that 
our three fits yield similar results and are all consistent with the high-quality BMW data, which are not included in our fits.

\section{Predicted baryon sigma terms} Using the best fit LECs, we predict the sigma terms of the octet baryons and tabulate the results in Table~\ref{Tab:N3LOsigma}.
Our predictions given by the LECs of Table~\ref{Tab:N3LOLECs} are consistent with each other within uncertainties, and the scale setting effects on the sigma terms seem to be small. Therefore, we take the central values from the fit to the mass independent $a$ fixed LQCD simulations as our final results, and treat the difference between different lattice scale settings as systematic uncertainties, which are given in the second parenthesis of the second column  of Table~\ref{Tab:N3LOsigma}.  It is clear that
for $\sigma_{\pi N}$, uncertainties due to scale setting are dominant, while for $\sigma_{sN}$ statistics errors are much larger, calling for improved LQCD simulations. It should be noted that we have studied the effects of virtual decuplet baryons and variation of the LECs $D$, $F$, $F_\phi$, and the renormalization scale $\mu$, and found that the induced uncertainties are negligible compared to those shown in Table~\ref{Tab:N3LOsigma}. Furthermore, as shown in Ref.~\cite{Ren:2013wxa}, continuum extrapolations have no visible effects on the predicted sigma terms.

\begin{figure}[t]
  \centering
  \includegraphics[width=8cm]{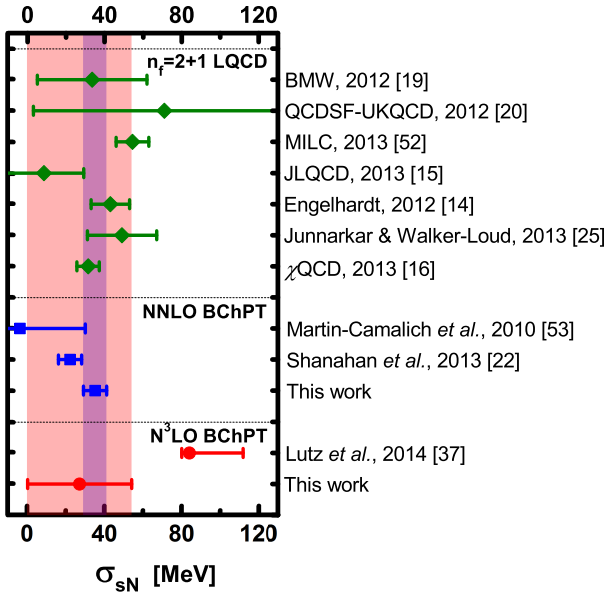}\\
  \caption{(color online). Strangeness-nucleon sigma term determined from different studies. The blue and red bands are
   our NNLO and N$^3$LO results, respectively. 
 }
  \label{Fig:sigma_s}
\end{figure}

The pion-nucleon sigma term, $\sigma_{\pi N}=55(1)(4)$ MeV, is in reasonable agreement with the latest $\pi N$ scattering study, $\sigma_{\pi N}=59(7)$ MeV~\cite{Alarcon:2011zs}, and also the systematic study of $n_f=2+1$ LQCD simulations on the nucleon mass, $\sigma_{\pi N}=52(3)(8)$ MeV~\cite{Alvarez-Ruso:2013fza}, but larger than that of Ref.~\cite{Lutz:2014oxa}, $\sigma_{\pi N}=39^{+2}_{-1}$ MeV. Our predicted $\sigma_{sN}$ is compared with those of earlier studies in Fig.~\ref{Fig:sigma_s}, classified into three groups according to the methods by which they are determined. The first group is the results reported by the $n_f=2+1$ LQCD simulations, while the second and third groups are  predicted by the NNLO and N$^3$LO BChPT, respectively. Our results are  consistent with the latest LQCD determinations and those of NNLO BChPT studies. It should be noted that, however, the prediction of the only other N$^3$LO study in the partial summation approach~\cite{Lutz:2014oxa} is not consistent with our result and most LQCD results.

A note of caution is in order. Clearly, using the spectrum method to determine the baryon sigma terms depends critically on the details of the LQCD simulations. Lattice scale setting is just one of the
sources for potentially large systematic errors. In the present work, we have studied three common alternative strategies and found that the resulting predictions remain almost the same.
Nevertheless, our studies do not exclude the possibility that predictions can change in more rare scenarios. In addition, other LQCD artifacts not addressed in the present work that  affect little the baryon masses may have an impact on the predicted baryon sigma terms, which is, however, beyond the scope of the present work.

\section{Convergence study}
Given the fact that BChPT plays an important role in predicting the baryon sigma terms, it is of particular importance to assess the uncertainties of truncating chiral expansions. It becomes even more important in the $u$, $d$, and $s$ three-flavor sector, where convergence is governed by the relative large ratio of $m_K/\Lambda_\mathrm{ChPT}\approx0.5$. Previous studies either stayed at NNLO or N$^3$LO and, therefore, were unable to perform such an analysis except those of Refs.~\cite{Ren:2012aj,Ren:2013dzt}, which, however, focused on a global study of the baryon masses and did not include all the QCDSF-UKQCD data that provide further constraints on the strange-quark mass dependence of the octet baryon masses.

To understand quantitatively the convergence issue, we have studied at NNLO the octet baryon masses of the PACS-CS, LHPC and QCDSF-UKQCD data obtained with the lattice spacings $a$ given by the LQCD collaborations themselves. We have allowed the LEC $F_\phi$ and the renormalization scale $\mu$ to vary to get an estimation of the induced variation.  All the obtained $\chi^2/\mathrm{d.o.f.}$ is larger than 1, indicating that higher-order chiral contributions need to be taken into account. In addition, we have employed
the FRR method, which is known to converge relatively faster, to study the same data and found no qualitative difference from the EOMS approach. We noted that  if one allows the $F_\phi$ to deviate from the chiral limit value to take into account SU(3) breaking effects, the so-obtained $F_\phi$ is close to its SU(3) average $1.17f_\pi$ with $f_\pi=92.1$ MeV~\cite{Beringer:1900zz}. The predicted strangeness-nucleon sigma term is shown in Fig.~\ref{Fig:sigma_s}.
It is clear  that the NNLO result has a much smaller uncertainty compared to the N$^3$LO one mainly because the LECs are over constrained by the LQCD simulations. It should be mentioned that in the Feynman-Hellmann method the large $m_s$ multiplying the derivative enhances the uncertainty in the determination of 
the strangeness-baryon sigma term, which seems to dominate the uncertainty and therefore puts an upper limit in the precision one can achieve.

\section{Conclusion}In this work,  we have performed a determination of the octet baryon sigma terms using the covariant baryon chiral perturbation theory up to next-to-next-to-next-to-leading order. We found $\sigma_{\pi N}=55(1)(4)$ MeV and $\sigma_{sN}=27(27)(4)$ MeV.
Special attention was paid to uncertainties induced by 
 the lattice scale setting method, which, however, were found to be small, in contrast with previous studies. Other uncertainties, such as  those induced by truncating chiral expansions and
 variations of  LECs were also studied in detail. In addition, we have used the strong-interaction isospin-splitting effects from the LQCD simulations to
 further constrain the relevant LECs.  Our results indicate a small scalar strangeness content in the nucleon, consistent with the strangeness contribution to the proton spin and to 
 the electromagnetic form factors of the nucleon.

\section{Acknowledgements}
We thank Dr. Jorge Martin-Camalich for a careful reading of this manuscript. X.-L.R acknowledges support from the Innovation Foundation of Beihang University for PhD Graduates.  This work was partly supported by the National Natural
Science Foundation of China under Grants No. 11005007, No. 11375024, and No. 11175002, the
New Century Excellent Talents in University Program of Ministry of Education of China under
Grant No. NCET-10-0029, the Fundamental Research Funds for the Central Universities, and the
Research Fund for the Doctoral Program of Higher Education under Grant No. 20110001110087.

\end{document}